\documentclass[%
 reprint,
 superscriptaddress,
 amsmath,amssymb,
 aps,
 prl,
]{revtex4-2}
\usepackage{siunitx}
\usepackage{xcolor}
\usepackage{lipsum} 

\usepackage{appendix}
\usepackage{braket}
\usepackage{float} 
\usepackage{chemformula}
\usepackage{graphicx}
\usepackage{dcolumn}
\usepackage{bm}
\usepackage{setspace}
\usepackage{mathtools}
\usepackage{braket}
\usepackage{graphicx}
\usepackage{dcolumn}
\usepackage{bm}
\usepackage[colorlinks=true, citecolor=blue, linkcolor=blue, urlcolor=blue]{hyperref}

\makeatletter
\newcommand*{\addFileDependency}[1]{
  \typeout{(#1)}
  \@addtofilelist{#1}
  \IfFileExists{#1}{}{\typeout{No file #1.}}
}
\makeatother

\setcitestyle{super}

\begin{document}

\title{Resolving self-cavity effects in two-dimensional quantum materials
}




\newcommand{\affiliationHavard}{
Lyman Laboratory, Department of Physics, Harvard University, Cambridge, MA 02138, USA
}

\newcommand{\affiliationDBSK}{Gwangju Institute of Science and Technology, 123 Cheomdangwagi-ro, Buk-gu, Kwangju, South Korea}

\newcommand{\affiliationRWTH}{
Institut f\"ur Theorie der Statistischen Physik, RWTH Aachen University and JARA-Fundamentals of Future Information Technology, 52056 Aachen, Germany
}
\newcommand{\affiliationMPSD}{
Max Planck Institute for the Structure and Dynamics of Matter,
Center for Free-Electron Laser Science (CFEL),
Luruper Chaussee 149, 22761 Hamburg, Germany
}

\newcommand{\affiliationBremen}{
Institute for Theoretical Physics and Bremen Center for Computational Materials Science,
University of Bremen, 28359 Bremen, Germany
}

\newcommand{\affiliationBristol}{
H H Wills Physics Laboratory, University of Bristol, Bristol BS8 1TL, United
Kingdom 
}

\newcommand{\CU}{Department of Physics, Columbia University,
538 West 120th Street, New York, NY 10027, USA}

\newcommand{\affiliationETH}{
Institute for Theoretical Physics, ETH Z\"urich, 8093 Z\"urich, Switzerland
}

\newcommand{\affiliationOxford}{
Department	of	Physics,	Clarendon	Laboratory,	University	of	Oxford,	United	Kingdom
}
\newcommand{\PKS}{Max Planck Institute for the Physics of Complex Systems, Nöthnitzer Straße 38, 01187 Dresden,
Germany}

\author{Marios H. Michael}
\thanks{Corresponding author: \href{mariosm@pks.mpg.de}{mariosm@pks.mpg.de}}
\affiliation{\PKS}
\affiliation{\affiliationMPSD} 
\author{Gunda Kipp}
\affiliation{\affiliationMPSD} 

\author{Alexander M. Potts}
\affiliation{\affiliationMPSD}
\affiliation{\CU}

\author{Matthew W. Day}
\affiliation{\affiliationMPSD}
\affiliation{\CU}

\author{Toru Matsuyama}
\affiliation{\affiliationMPSD}

\author{Guido Meier}
\affiliation{\affiliationMPSD}

\author{Hope M. Bretscher}
\thanks{Corresponding author: \href{hope.bretscher@mpsd.mpg.de}{hope.bretscher@mpsd.mpg.de}}
\affiliation{\affiliationMPSD}
\affiliation{\CU}
\author{James W. McIver}
\thanks{Corresponding author: \href{jm5382@columbia.edu}{jm5382@columbia.edu}}
\affiliation{\affiliationMPSD}
\affiliation{\CU}

\date{\today}

\begin{abstract}
Two-dimensional materials and van der Waals (vdW) heterostructures host many strongly correlated and topological quantum phases on the $\sim$ meV energy scale. Direct electrodynamical signatures of such states are thus expected to appear in the terahertz (THz) frequency range (1 THz $\sim$ 4 meV). Because the typical size of vdW heterostructures ($\sim$10 $\mu m$) is much smaller than the diffraction limit of THz light, probing THz optical conductivities necessitates the use of near-field optical probes. However, interpreting the response of such near-field probes is complicated by finite-size effects, the presence of electrostatic gates, and the influence of the probe itself on material dynamics --- all of which conspire to form polaritonic self-cavities, in which interactions between THz electromagnetic fields and material excitations form discretized standing waves. In this paper, we demonstrate the relevance of self-cavity effects in 2D materials and derive an analytical framework to resolve these effects using the emerging experimental technique of time-domain on-chip THz spectroscopy. We show that by pairing experiments with the analytical theory, it is possible to extract the THz conductivity and resolve collective mode dynamics far outside the light cone, with $\sim \mu m$ in-plane and $\sim nm$ out-of-plane resolution. This study lays the groundwork for studying quantum phases and cavity effects in vdW heterostructures and 2D quantum materials. 
\end{abstract}
\maketitle


\paragraph{Introduction.} Constructing heterostructures with atomic precision layer by layer using van der Waals (vdW) materials continues to fuel the discovery of novel quantum phases \cite{kennes2021moire,ren20252d,nuckolls2024microscopic}.
Two-dimensional crystals with different orientations or lattice constants can be stacked to create moiré heterostructures with flat bands and interaction-driven physics, giving rise to a rich array of correlated phenomena \cite{mak2022semiconductor,balents2020superconductivity,nuckolls2024microscopic}. The low density of states in these 2D structures means that their chemical potential and electronic band structure can be substantively tuned \textit{in situ} using electrostatic gates. This design and \textit{in situ} control over vdW heterostructures make them an ideal platform for investigating quantum phases, while their small size offers promising pathways for next-generation chip-based technologies\cite{ren20252d}. 

Many quantum phases and novel phenomena found in these materials have fingerprints on the $\sim \rm meV$ energy / $\rm THz$ frequency scale. For instance, the bandwidth of moir\'e flatbands often spans a few $\rm meV$\cite{Bistritzer11}, as do energy gaps associated with phenomena such as superconductivity \cite{jindal2023coupled,balents2020superconductivity,cao2018unconventional,xia2025superconductivity,guo2025superconductivity}, Wigner crystallization\cite{Tomasz25}, charge ordering\cite{Pavel20}, and correlated magnetic insulators\cite{park2023observation}. In addition, exotic collective modes in correlated phases, such as Cooper pair plasmons and Higgs modes in two-dimensional superconductors \cite{sun2020collective}, phasons in excitonic insulators\cite{Edoardo23,bretscher2021imaging}, and magnons\cite{Rongione23} in magnetically ordered systems, have also been reported and predicted to occur at THz frequencies.

\begin{figure*}[t!]
    \centering
    \includegraphics[width=0.9 \linewidth]{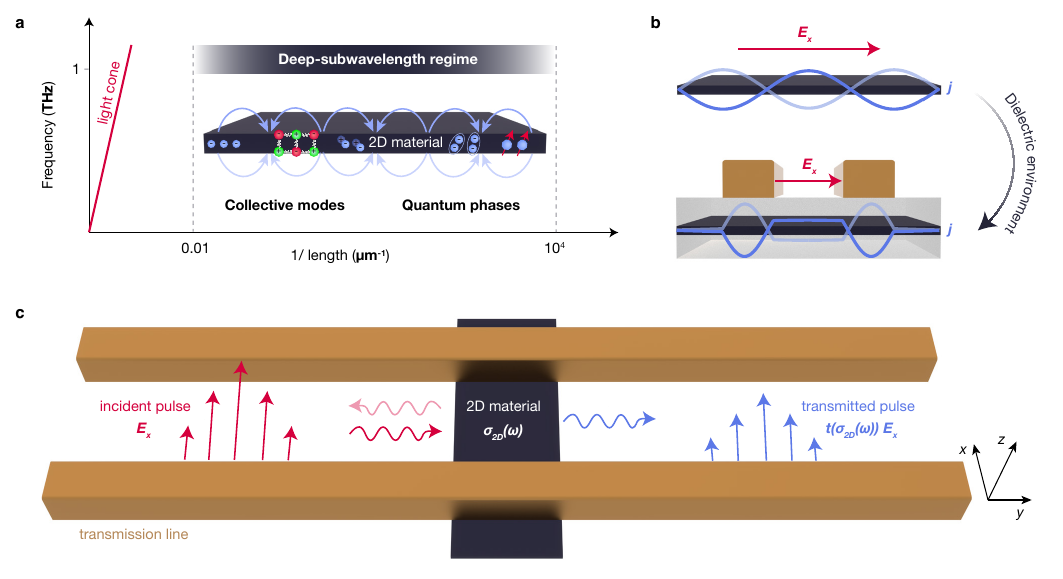}
     \caption{\textbf{Self-cavity effects in on-chip THz spectroscopy: a} Many of the emergent quantum phases and collective modes in 2D materials within vdW heterostructures manifest at THz frequencies, placing these systems -  relative to typical device dimensions - deep within the subwavelength regime, far below the light cone. \textbf{b} Owing to their subwavelength dimensions, 2D materials act as self-cavities for THz light ($E_x$), with their edges imposing sharp boundaries for finite-momentum current excitations ($j$). In addition to the finite width of the sample, the surrounding dielectric environment, especially proximitized near-field probes, such as the metal strips of on-chip THz spectroscopy circuits, plays a crucial role in shaping the current distribution of a self-cavity mode. \textbf{c} Principle of on-chip THz spectroscopy: An incident THz pulse with electric field $E_x$ propagates along a transmission line, consisting of coplanar striplines, toward an integrated 2D material characterized by a bare conductivity $\sigma_{\text{2D}}(\omega)$. Upon reaching the 2D material, part of the pulse is reflected (red) and part is transmitted (blue), as the electric field excites self-cavity modes within the material, as illustrated in (b). By detecting the conductivity-dependent transmitted pulse $t(\sigma_{\text{2D}}(\omega))$, and carefully accounting for the influence of the metal strips and surrounding dielectric environment, the bare conductivity $\sigma_{\text{2D}}$ can be accurately extracted with the help of the analytical theory presented in this work.}
    \label{fig:Sketch}
\end{figure*}

In the race to uncover novel physics enabled by atomically thin heterostructures, optical probes resonant with the characteristic energy scales of the emergent phenomena are essential\cite{armitage2009electrodynamics,basov2011electrodynamics}. Yet, a key challenge arises: how can the THz conductivity of small samples, just a few nanometers thick and micrometers in lateral size, be accurately probed? As shown in Fig.~\ref{fig:Sketch}(a), these dimensions are small compared to the wavelength of THz light ($\sim 300~\mu\mathrm{m}$ at 1 THz), placing vdW heterostructures deep in the sub-wavelength regime, where well-established far-field methods used to extract the frequency-dependent conductivity, $\sigma(\omega)$\cite{koch2023}, fail. Instead, near-field probes are needed, where THz radiation is confined and funneled to a small sub-wavelength region comparable to the size of the sample. Subwavelength focusing of THz light has been achieved, for instance, using sharp metal tips in THz Scanning Near field Optical Microscopy (SNOM)\cite{hillenbrand2025visible,cocker2021nanoscale,huber2008terahertz,mastel2017terahertz,haeuser2024analysis,xu2023electronic}, local THz emitters in THz microscopy \cite{AlexMarios, handa20242d}, and the method explored in this article, metallic coplanar striplines in on-chip THz spectroscopy \cite{kipp2024cavity,zhao2023observation, gallagher2019quantum, seo2024onchipterahertzspectroscopydualgated,potts2023chip,chen2024directmeasurementterahertzconductivity,island2020chip,zhong2008terahertz,sterbentz2023chip,sprik1987far} (See Fig.~\ref{fig:Sketch} (c)).

However, extracting the complex optical conductivity or analogous material response functions from near-field probes has, to date, remained challenging. This is because the small size of vdW flakes not only limits far-field optical methods, but makes edge reflections of collective modes a dominant effect in their optical response\cite{kipp2024cavity}. The sample's edges force excited currents to form standing waves, such that the sample itself acts as a near-field \textit{self-cavity} for THz light. These optically active standing waves are highly sensitive to the surrounding environment, meaning that any THz near-field probe can significantly alter the material’s optical response \cite{hillenbrand2025visible,fei2012gate} and shape the effective self-cavities formed by the edges, as shown schematically in Fig.~\ref{fig:Sketch} (b). For example, reflections of THz plasmons from sample edges have been used as a feature in THz-SNOM to study the velocity of charge propagation at a given frequency \cite{lundeberg2017tuning, xu2023electronic}. Similarly, the THz spintronics emitter used in THz microscopy has been shown to strongly influence the effective conductivity of the target material\cite{AlexMarios}.

The above discussion underscores a fundamental requirement: to accurately extract the complex THz conductivity of vdW heterostructures, any near-field probe must be paired with an analytical theory that links the optical response to the bare conductivity, decoupling edge effects and probe influences. Without this direct mapping, inferring conductivity becomes impractical, requiring costly 3D simulations for an effectively infinite range of sample conductivities, probing geometries, and heterostructure dimensions \cite{hillenbrand2025visible}. 

In this paper, we deliver an analytical mapping between the frequency-dependent complex cavity conductivity measured using on-chip THz spectroscopy, and the bare conductivity, positioning this technique as the only method capable of resolving these responses of vdW heterostructures within the THz frequency range. Furthermore, we demonstrate that self-cavity effects can be intentionally controlled through the stripline geometry to selectively probe finite momentum excitations with $\sim \mu m$ resolution in-plane and $\sim nm$ to $\textup{\AA}$ resolution along the out-of-plane, $z$-direction (See Fig.~\ref{fig:Sketch} (c)). Specifically, we show how on-chip THz spectroscopy can reveal optically silent modes, such as demon and shear optical phonon modes. Furthermore, we explore how strongly correlated electron phenomena, such as 2D superconducting gaps, correlated insulating gaps, strange metal plasmons, and exciton polaritons, manifest in the cavity conductivity spectrum. These self-cavity-dominated spectra can uncover material characteristics that would be hidden in the bare conductivity accessed in far-field probes, making on-chip THz spectroscopy, in combination with the analytical theory presented in this work, a powerful tool for exploring a wide range of emergent phases and collective modes.

\paragraph{Extracting the THz 2D conductivity.}
\label{sec:Single}
\begin{figure*}[t!]
    \centering
    \includegraphics[width=0.9\linewidth]{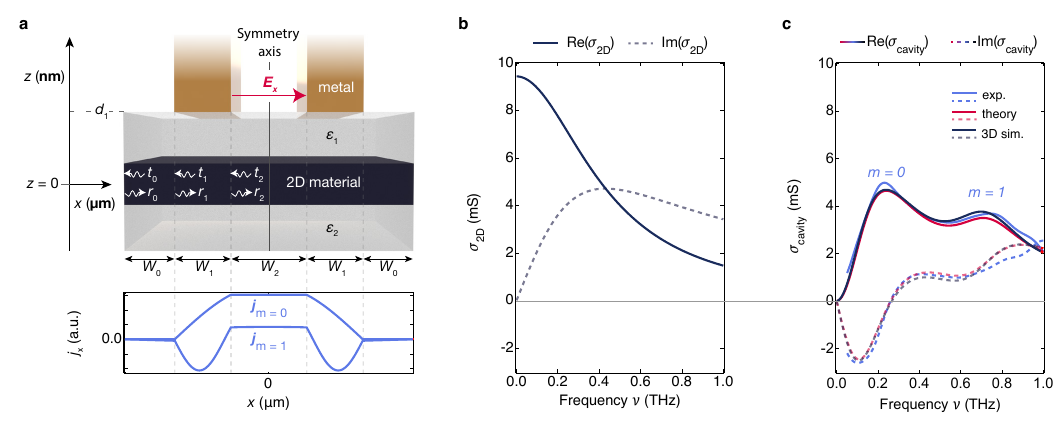}
    \caption{\textbf{Cavity conductivity of a single layer: a} Top panel: Cross-section perpendicular to the propagation of the stripline mode. The electric field from the probe pulse excites polaritonic standing waves, while the presence of the metal strips split the region into screened regions below the strips and unscreened regions outside. Bottom panel: Sketch of the current density as a function of position in the material, at the frequencies corresponding to the two plasmon polariton peaks in (c). \textbf{b} Real and imaginary part of graphite thin film 2D conductivity, $\sigma_{\text{2D}} = \frac{\epsilon_0 \omega_{pl}^2 d_{gr}}{- i \omega + \gamma}$, with $\omega_{pl} = 2 \pi \times 121$ THz, $d_{\text{gr}}=$~\SI{5}{nm}, $\gamma =$~\SI{0.43}{THz} and $\omega$ the angular frequency. Corresponding cavity conductivity measured by the on-chip response for $W_0=$~\SI{1}{\micro\meter}, $W_1=$~\SI{3.33}{\micro\meter}, $W_2=$~\SI{2.56}{\micro\meter}, $d_{\text{hBN}}=$~\SI{24}{nm} and $C_{\text{filling}}$~=~~\SI{319}{mm^{-1}}. The 2D material was separated from the metal strips by thin hBN with $\epsilon_{1} =  3.7 $ placed on top of sapphire, $\epsilon_2 = 10$. The plot shows the results of experiments, 3D numerical simulations and the analytical model. The agreement between the three curves provides strong evidence for the validity of the analytical model. Here and for the rest of the paper we plot frequency units where $\omega = 2\pi \nu$. 
    }  \label{fig:2Dconductivity}
\end{figure*}

The on-chip THz spectroscopy circuitry is sketched in Fig.~\ref{fig:Sketch} (c). Photoconductive switches \cite{auston1975picosecond}  triggered by femtosecond optical pulses are used to launch, and read out, a THz pulse, from which the frequency dependent transmission coefficient can be extracted. Specifically, the first switch launches a \textit{quasi}-TEM mode into the gold coplanar striplines which propagates along the $y$-direction, while the electric field lines are oriented along the $x$-direction between the metal strips. Once the pulse reaches the 2D material, the electric field excites 2D currents along the $x$-direction that influence the dispersion of the stripline mode from which the transmission coefficient is derived. In the thin film along the $y$-direction limit, the coefficient is approximately given by a simple lumped model approximation:

\begin{equation}
    t = 1-\frac{Y_{\text{lumped}} Z_{\text{0}}}{2},
\end{equation}

\noindent
where $t$ is derived as the ratio between the transmitted voltage across the coplanar striplines to a reference field that was transmitted in the absence of the material\cite{kipp2024cavity}. $Z_{\text{0}} =  \left(\sqrt{\epsilon_{\text{sl}}}\epsilon_0 c\right)^{-1}$ is the impedance of the striplines and $Y_{\text{lumped}}$ the effective admittance of the lumped model, which is proportional to the sum of 2D conductivities of different layers in the target heterostructure: 
\begin{equation}
    Y_{\text{lumped}} = L_y \sum_i \frac{1}{H_{\text{eff}}} \sigma_{\text{i, cavity}},
    \label{eq:Lumped}
\end{equation}
where $L_y$ is the dimension of a heterostructure along the $y$-direction, $i$ labels the different elements of a heterostructure and $\sigma_{i,cavity}$ is the effective 2D conductivity of each layer experienced by the THz pulse. For a single 2D material, there is only one layer, but usually such a system is also encapsulated in an insulator such as a layer of hBN. $H_{\text{eff}}$ is an overall proportionality constant that depends on the stripline mode structure and is independent of the target material as shown in Suppl. Note 1\cite{Supplemental}. The derivation of the transmission coefficient and corrections beyond the thin-film approximation are presented in Suppl. Note 1\cite{Supplemental}. 

Equation~(\ref{eq:Lumped}) shows that the effective cavity conductivity, $\sigma_{cavity}$ is readily available in transmission, however the challenge comes from what was discussed in the introduction; the cavity conductivity is strongly affected by the edges of the flake and the presence of the striplines. As shown in Fig.~\ref{fig:2Dconductivity}(a), the electric field in between the striplines excites 2D polaritonic standing waves that bounce back and forth in the material. The current distribution is given by:
\begin{widetext}
\begin{equation}
    j_{\text{2D}}(x,t) = e^{i\omega t} \left\{ \begin{matrix*}[l] t_0 E_{\text{ext}} e^{i q_{\text{un}} x}  + r_0 E_{\text{ext}} e^{ - i q_{\text{un}} x} ,&&
    - W_0 - W_1 - \frac{W_2}{2} < x < - W_1 - \frac{W_2}{2} \\
    t_1 E_{\text{ext}} e^{i q_{\text{sc}} x}  + r_1 E_{\text{ext}} e^{ - i q_{\text{sc}} x} ,&&
    - W_1 - \frac{W_2}{2} < x < - \frac{W_2}{2} ,\\
     t_2 E_{\text{ext}} e^{i q_{\text{un}} x}  + r_2 E_{\text{ext}} e^{ - i q_{\text{un}} x} + \sigma_{\text{2D}}(\omega) E_{\text{ext}},&& - \frac{W_2}{2} < x < + \frac{W_2}{2}
    \end{matrix*} \right.,
\end{equation}
\end{widetext}
while the current density in the other regions of the flake is determined by symmetry. In region 2, the middle of the striplines, a current is directly excited through the term $\sigma(
\omega) E_{\text{ext}}$. However, the current density needs to also satisfy boundary conditions at the sample edges and between screened and unscreened regions at the edge of the striplines \cite{jiang2018theory}. This requirement sets up polaritonic standing waves, with a different momentum in each region determined by the screened and unscreened polariton dispersion. To a very good approximation the dispersion of screened and unscreened polaritons for a general conductivity is given by the formulas:
\begin{equation}
    q_{\text{sc}} = \sqrt{\frac{ i \omega \epsilon_1 \epsilon_0}{ d_1 \sigma_{\text{2D}}(\omega)}}, \mbox{\qquad} q_{\text{un}} = \frac{i \omega(\epsilon_1 + \epsilon_2)\epsilon_0}{\sigma_{\text{2D}}(\omega)},
\end{equation}
where $\sigma_{\text{2D}}(\omega)$ is the bare conductivity of a single layer, $d_1$ is the distance of the 2D layer to the stripline and $\epsilon_1$ and $\epsilon_2$ are the permittivities of insulating materials surrounding the 2D material. In Suppl. 2\cite{Supplemental}, we comment on the accuracy of these intuitive expressions, show how they can systematically be made more accurate, and demonstrate that these formulas can be generalized for the case of an anisotropic dielectric like the commonly-used hBN. 

The cavity conductivity then corresponds to the self-consistently determined average current response in region 2, i.e. $\sigma_{cavity}(\omega)= \frac{1}{W_2} \frac{\int_{-W_2/2}^{W_2/2} d x j_{\text{2D}}(x)}{E_{\text{ext}}}$. In all the examples treated in this article, the unscreened momentum is much smaller than the screened and this expression simplifies to $\sigma_{cavity} = \sigma_{\text{2D}} + r_2 + t_2 $. The factor $r_2 + t_2$ captures the deviation of the cavity conductivity from the bare conductivity, representing the feedback response from the edges of the sample and boundary reflections between screened and unscreened regions. One of the main results of this paper is the derivation of an analytical formula for the cavity conductivity for the structure shown in Fig.~2 (a). The analytical formula is presented in Suppl. Note 2\cite{Supplemental} and takes the form: \begin{equation}
    \sigma_{cavity} = \sigma_{\text{2D}} (\omega)\bigg( 1 +
    f\left( \{W_i\}, q_{\text{sc}}(\omega), q_{\text{un}}(\omega) \right) \bigg),
    \label{eq:CondSingle}
\end{equation}
where $f$ is the feedback which explicitly depends on the geometry, $\{W_i\} = \{W_0, W_1, W_2\} $. Equation~(\ref{eq:CondSingle}) shows that the effective conductivity depends on both the geometry and the polaritonic response of the material. As an example, we consider a 2D metal with Drude bare conductivity, shown in Fig.~2(b). The corresponding cavity conductivity, derived from our analytical theory is shown in Fig.~2(c). A notable difference between the bare and cavity conductivities emerges in the limit $\omega \rightarrow 0$: while the bare conductivity retains its Drude behavior, the cavity conductivity vanishes. This is a general feature of the cavity response, resulting from the fact that no DC current can flow between the striplines and the material. Another key feature is the presence of resonances in the cavity conductivity. These resonances are dominated by standing waves in the screened region, as illustrated by the current distribution in Fig.~\ref{fig:2Dconductivity}(a), and correspond approximately to quarter- and three-quarter-wavelength modes. In contrast, the current in the unscreened region remains nearly constant, since $q_{\text{un}} \ll q_{\text{sc}}$. The rate at which the cavity conductivity approaches zero at low frequencies together with the appearance of plasma resonances at higher frequencies are governed entirely by the polaritonic response and finite-momentum plasmon-polariton dispersion. Thus, in the low-frequency regime--relevant for 2D correlated materials--geometric effects and polaritonic phenomena play a dominant role.

The above result underscores the potential of on-chip THz spectroscopy: 1) it can extract the bare complex conductivity response, 2) it directly probes finite momentum THz collective mode dynamics with micrometer sensitivity and 3) the geometry can be purposefully designed to shape the light-induced response of the material. In the realm of spectroscopy, this geometric control can be leveraged to enhance sensitivity and contrast of specific aspects of the material response. 

The example chosen for Fig. 2 is that of a microstructured graphite flake, similar to those used as electrostatic gates in vdW heterostructures, for which we have both experimental and numerical data. In Fig.~\ref{fig:2Dconductivity} (b) we show the bare conductivity of graphite, as captured by the Drude model, while in Fig.~\ref{fig:2Dconductivity} (c) we show the cavity response. The experimental, numerical and analytical cavity response are in remarkable agreement with each other. This agreement demonstrates that the analytical theory  captures the dominant contributions to the full solution. Discrepancies between the numerical and analytical solutions at higher frequencies are attributed to coupling to far-field photons explicitly excluded by the analytical theory. Such discrepancies are minimal, and support our deep sub-wavelength approximation that ignores radiative corrections. Further discussion on discrepancies between the full 3D simulations and the analytical model is presented in Suppl. Note 3\cite{Supplemental}. Details of how the experimental data were acquired are discussed in Suppl. Note 4\cite{Supplemental}.





\paragraph{Bilayer and gated 2D materials.}
\label{sec:bilayer}
\begin{figure}
    \centering
    \includegraphics[width = 1. \linewidth]{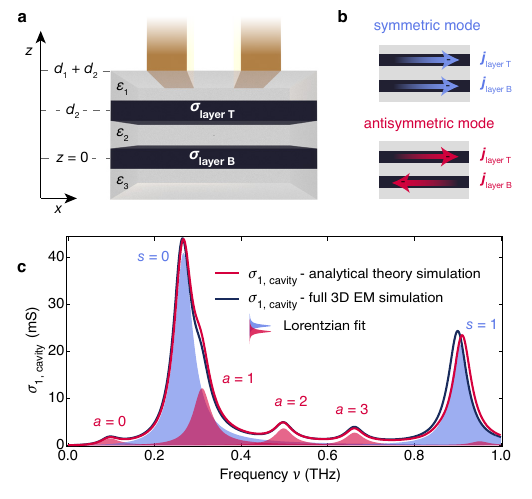}    \caption{\textbf{Probing the cavity conductivity of multilayer and gated 2D materials: a} Cross-section of a heterostructure consisting of a top and bottom layer with bare 2D conductivities $\sigma_{\text{layer T}}$ and $\sigma_{\text{layer B}}$ embedded in metallic coplanar striplines. \textbf{b} Currents that form in the top and bottom layer hybridize to form in-phase (``symmetric mode") and out-of-phase (``antisymmetric mode") oscillations of current density. \textbf{c} Theoretical calculations of the cavity conductivity of the heterostructure shown in (a) using the analytical theory and full 3D electromagnetic simulations. The largest amplitude resonances (shaded in blue and labeled with $s=0,1$) correspond to symmetric modes and the lower amplitude resonances are formed by antisymmetric modes (shaded in red and labeled with $a=0,1,2,3$). Parameters: $W_0=$~\SI{1}{\micro\meter}, $W_1=$~\SI{3.5}{\micro\meter}, $W_2=$~\SI{2.5}{\micro\meter}, $d_{\text{1}}=$~\SI{20}{nm}, $d_{\text{2}}=$~\SI{20}{nm}, where each layer consists of identical metal films, $\sigma_{T} = \sigma_B = \sigma_{2D}$. The Drude bare conductivity is $\sigma_{\text{2D}} = \frac{\epsilon_0 \omega_{pl}^2d}{- i \omega + \gamma}$, with $d =$~\SI{4}{nm}, $\omega_{pl} = 2 \pi \times 118$ THz and $\gamma = 0.43$ THz. For the numerical simulation we used a length along the $y$-direction of $L_y=$~\SI{7}{\micro\meter}. For all systems, it was assumed that the 2D material was encapsulated by thin $hBN$ so that $\epsilon_{1} = \epsilon_2 =  3.7 $ and placed on top of sapphire, $\epsilon_3 = 10$.}
    \label{fig:Bilayer}
\end{figure}

One advantage in 2D vdW materials is their \textit{in situ} chemical potential and displacement field tunability. This control is typically achieved by applying a voltage between a 2D metal and the sample of interest, separated by a thin dielectric spacer.  
A sketch of this scenario is presented in Fig.~\ref{fig:Bilayer} (a) where the vdW material is captured through $\sigma_{layer\ T}$ and the metallic gate is captured through $\sigma_{layer\ B}$. As discussed in the previous section, the plasmonic properties of semi-metallic flakes commonly utilized as electrostatic gates lead to a non-trivial THz conductivity. Indeed,  Fig.~\ref{fig:Bilayer}(b) shows that the plasmonic responses of a gate and vdW material hybridize into symmetric and anti-symmetric polaritons, which also have screened and unscreened counterparts depending on whether they lie underneath the gold strips or not.  As an example, we present here the approximate analytical formula for the momenta in the unscreened region:
\begin{align}
    q_{un,a} = \sqrt{\frac{i \omega \epsilon_2 \left( \sigma_{\text{T}} + \sigma_{\text{B}} \right) \epsilon_0 }{d_1 \sigma_{\text{T}}  \sigma_{\text{B}}}} , \mbox{\quad} q_{un, s} = \frac{ i \omega \left( \epsilon_1 + \epsilon_3 \right) \epsilon_0 }{\sigma_{\text{T}} + \sigma_{\text{B}}},
\end{align}
where the symmetric mode corresponds to the 2D plasmon of the combined vdW material and gate, while the anti-symmetric mode can be thought off as the plasmon of the low conductive material screened by the high conductive material\cite{sarma1981collective}. This is seen by the fact that if $\sigma_B >> \sigma_T$, the anti-symmetric unscreened mode is reduced to a screened mode of layer one. These expressions are analytical but approximate and can be systematically improved through Newton's method which we present in the Suppl. Note 2\cite{Supplemental}. Expressions for the screened bilayer modes are also included in the Suppl. Note 2\cite{Supplemental}. 

The conductivity experienced by the stripline mode is $\sigma_{cavity} = \sigma_T(\omega) (1 + f_T(\omega)) + \sigma_B(\omega)(1 + f_B(\omega))$, where $f_T$ and $f_B$ correspond to the feedback from the boundaries and the screened region that is found through solving boundary conditions on the two layers. Complete expressions are presented in the Suppl. Note 2\cite{Supplemental}. As in the previous section, the screened region acts as a cavity resonator, and the cavity conductivity has signatures of peaks originating from both the symmetric and anti-symmetric modes. 

In Fig.~\ref{fig:Bilayer} (c), we plot as an example the case of two identical thin metals. As before, we compare our analytical theory with simulations and find remarkable agreement. Another relevant example of doped graphene on top of thin graphite, which acts as the electrostatic gate, has been addressed in a sister experimental publication by the authors~\cite{kipp2024cavity}. There, a complete characterization of the ultra-strong coupling between plasmonic effects of the graphite gate and the graphene Dirac plasmon is presented.

\begin{figure}
    \centering
    \includegraphics[width = 1. \linewidth]{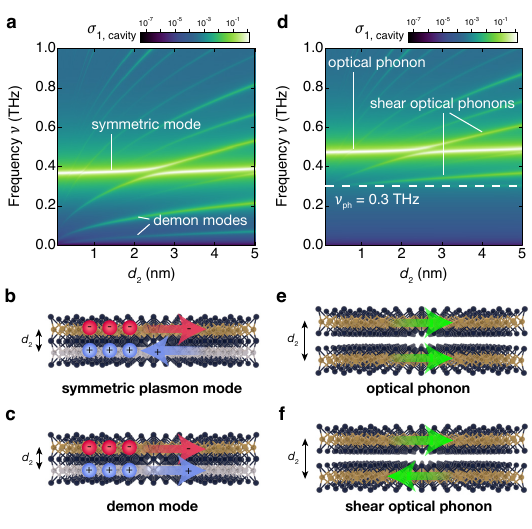}
    \caption{\textbf{Detecting demon modes and phonons with Angstrom precision: a} Cavity conductivity spectrum of a single layer with two bands separated in the z-direction by $d_2$. Demon modes and symmetric plasmon modes are detected.\textbf{b} The usual 2D plasmon corresponds to out-of-phase oscillation of electrons and holes. \textbf{c} A demon mode is the collective mode where electrons and holes oscillate in-phase. \textbf{d} Cavity conductivity spectrum of a bilayer with optically active phonon modes.\textbf{e} If adjacent layers of the crystal lattice move in the same directions, an optical phonon is excited \cite{pizzi2021shear} .\textbf{f} If adjacent layers of the crystal lattice move in opposite directions, shear optical phonon is excited which in the far-field is optically silent\cite{pizzi2021shear}. Parameters given in Suppl. Note 5\cite{Supplemental}.}
    \label{fig:Demons}
\end{figure}

\paragraph{Optically silent mode spectroscopy.}
\label{sec:OptSil}
An interesting point to note is that the graphene-like resonances in the experiment of reference\cite{kipp2024cavity} already correspond to an optically silent acoustic plasmon, since the graphene is screened by graphite and the total current in the two layers is zero. This demonstrates the power of on-chip THz spectroscopy to probe optically silent modes that consist of dipoles separated by $\sim nm $ or $\textup{\AA}$s along the $z$ - direction. This sensitivity is a consequence of the nanometer proximity of the gold strips to the heterostructure which couples more strongly to the top layer than the bottom. Far-field optical probes are sensitive to the net-dipole moment of a system, and thus anti-symmetric oscillations are invisible, or optically silent, using traditional linear spectroscopic probes \cite{husain2023pines}. Such excitations can be found in many contexts, such as phase modes of bilayer Wigner Crystals\cite{Tomasz25}, anti-symmetric modes of bilayer quantum wells\cite{sarma1981collective} and moir\'e phonon modes \cite{ramos2025flat}. In this section, we focus on two specific examples, demon modes and bilayer shear optical phonons presented in Fig.~\ref{fig:Demons}, with detailed calculations presented in the Suppl. Note 5\cite{Supplemental}.

A demon mode corresponds to the in-phase oscillation of electron and hole carriers from different electronic bands, or equivalently the out-of-phase oscillations of their current\cite{pines1956electron,husain2023pines}, as shown in Fig.~\ref{fig:Demons} (c). The typical 2D plasmon corresponds to out-of-phase oscillations of electron and hole bands shown in Fig.~\ref{fig:Demons} (b). If the orbitals of each band are spatially separated in the z - direction then they can be detected by on-chip THz spectroscopy. As we show in Fig.~\ref{fig:Demons} (a) the larger the spatial distance along the $z$ direction between the bands, the more pronounced the demon mode resonances appear. On-chip THz signatures of demon modes could illuminate the underlying band structure of bilayer materials\cite{seiler2022quantum,jindal2023coupled,fei2018ferroelectric}, the origin of out-of-plane ferroelectricity \cite{jindal2023coupled,fei2018ferroelectric,zheng2020unconventional}, or the nature of correlated insulating energy gaps observed across a range of vdW bilayer systems \cite{wong2020cascade,park2023observation,nuckolls2024microscopic,qi2025perfect,nguyen2025perfect}.

Shear optical phonon modes involving the anti-symmetric oscillation of two dipole active phonons in each layer of a bilayer are sketched in Fig.~\ref{fig:Demons}(f). These excitations are also optically silent using far-field IR-active probes but can become bright in the near-field response, as shown in Fig.~\ref{fig:Demons} (d). The symmetric optical phonon excitation (sketched in Fig.~\ref{fig:Demons}(e)) appears as a bright signature, while the shear optical phonons appear as side peaks. Signatures of such phonon modes can help identify the microscopic order underlying many correlated phases in vdW materials \cite{ramos2025flat,papaj2023probing}.  

Similar to the gated material case, while the symmetric contribution corresponds to the largest peak, side peaks are observed which originate from optically silent demon modes and shear optical phonon modes and the geometry of the spectroscopy can be leveraged to extract the dispersion of these modes. We note that these modes are polaritons and their dispersion is dominated by light.

\paragraph{Correlated 2D electron systems.}
\begin{figure*}[t!]
 \centering
    \includegraphics[width = 1. \linewidth]{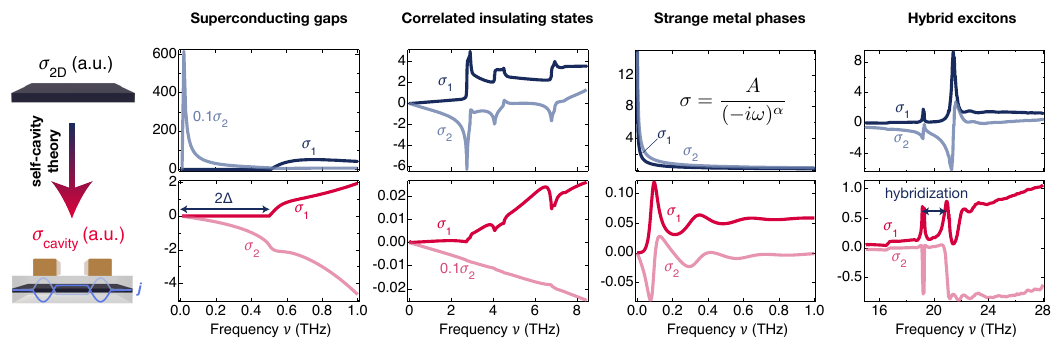}
    \caption{\textbf{On-chip THz spectroscopy of correlated materials:} The analytical cavity theory of this work can be used to simulate how the THz cavity conductivity spectrum looks like (red) for a given bare 2D conductivity of a phenomenon or collective mode in a 2D quantum material (blue), such as for a superconductor, a correlated insulating state, a strange metal phase or for hybrid excitons. The real and imaginary parts of the input 2D conductivities (blue) were taken from measured or simulated data. Superconducting conductivity was computed using the Zimmermann model (BCS theory). 
    The input curves for the correlated insulating states were extracted from measured data on twisted bilayer graphene (Ref.~\cite{Calderon2020TBG}), using the following cavity geometry: $W_0=$~\SI{3}{\micro\meter}, $W_1=$~\SI{3}{\micro\meter}, $W_2=$~\SI{3}{\micro\meter} and $d_{\text{1}}=$~\SI{24}{nm}. The input curves for the strange metal phases were simulated with the formula provided in Ref.~\cite{antoinegeorgesstrangemetalref} with $\alpha=0.7$, and in the following cavity geometry: $W_0=$~\SI{0}{\micro\meter}, $W_1=$~\SI{20.33}{\micro\meter}, $W_2=$~\SI{2.56}{\micro\meter} and $d_{\text{1}}=$~\SI{24}{nm}. 
    The input data for the hybrid excitons was extracted from measured data on bilayer graphene (Ref.~\cite{doi:10.1126/science.aam9175}) using the following cavity dimensions: $W_0=$~\SI{3}{\micro\meter}, $W_1=$~\SI{3}{\micro\meter}, $W_2=$~\SI{3}{\micro\meter} and $d_{\text{1}}=$~\SI{24}{nm}. For all systems, it was assumed that the 2D material was separated from the metal strips by thin $\rm hBN$ so that $\epsilon_{1} = 3.7 $, and placed on top of sapphire, $\epsilon_2 = 10$.}
    \label{fig:Gaps}
\end{figure*}

In this section, after having established the theoretical framework, we show how on-chip THz spectroscopy can be used to probe finite momentum polaritons as well as extract the bare THz conductivity of a collection of different correlated electronic states. In Fig.~\ref{fig:Gaps}, the top panel in blue corresponds to the bare 2D conductivity, while in red, the self-cavity 2D conductivity is shown for different geometries that are specified in the caption. 

The self-cavity response of a superconductor exhibits a gap in $\sigma_1$ at $2\Delta$, equal to that of the bare conductivity response. Absent from the self-cavity imaginary response is the characteristic $1/\omega$ scaling at low frequencies observed in the blue, bare 2D conductivity. Instead, $\sigma_{2,\text{cavity}}$ grows with increasing frequency, with a pronounced kink near $2\Delta$. 

We note that to observe such features, the overlap of the superconductor with the screened region is intentionally minimized to eliminate the competing superconducting plasmon response. This example shows that on-chip spectroscopy could be used to extract the THz superconducting gaps in unconventional moir\'e superconductivity\cite{MoireSC}.

On-chip THz spectroscopy can also be used to quantitatively extract the magnitude of insulating gaps. As we show in the example in Fig.~\ref{fig:Gaps}, the gap can be extracted through the real part of the cavity conductivity. Moreover, other sharp features in the conductivity, such as van Hove singularities, correspond to discontinuities in the gradient of the on-chip THz response. Here correlated insulator is used as a catch-all term to describe a range of different systems such as Mott insulators\cite{regan2020mott}, Wigner crystals\cite{regan2020mott}, ferromagnets\cite{zhou2022isospin,huang2018electrical}, anti-ferromagnets\cite{kang2020coherent,mai2021magnon} and ferroelectrics,\cite{jindal2023coupled,fei2018ferroelectric,zheng2020unconventional} all of which exhibit insulating behavior with a gap. Correlated insulating states have been observed in many flat-band systems\cite{huang2021correlated}. However, due to the high degree of degeneracy, particularly in twisted graphene structures, the microscopic origins of resulting symmetry-broken phases have remained a topic of ongoing investigation\cite{zondiner2020cascade}. Spectroscopically measuring gap magnitudes in encapsulated structures as a function of tuning parameter, which provides much finer resolution than that achievable with thermally activated transport, can constrain or identify the nature of symmetry breaking \cite{basov2011electrodynamics}.

The third example we consider is the case of strange metals. Strange metals are found on many of the canonical strongly-correlated phase diagrams near quantum phase transitions, and their relation to often proximate unconventional superconducting states remains a long-open question in condensed matter \cite{alexandradinata2020future}. These phases are characterized by the absence of well-defined electronic quasiparticles which leads to anomalous scattering rates, self-similar behavior (scale invariance), and long-range entanglement. Linear-in-T resistivity at low temperatures is one signature consistent with strange metallicity and has been observed in a number of vdW systems\cite{cao2020strange, jindal2023coupled,xia2025superconductivity}. In on-chip THz spectroscopy, at frequencies higher than the temperature, the self-similar and fractal behavior of the conductivity can be directly probed, which follows a power law behavior, $\sigma(\omega) = \frac{A}{\left( - i \omega \right)^\alpha}$, with a fractional power $\alpha$\cite{michon2023reconciling}. Such an expression has no intrinsic length scale and it gives rise to strange plasmons that have a fractal dispersion relation, $q_{\text{sc}} = \frac{i (- i \omega)^{\frac{\alpha +1}{2}} }{\sqrt{d_1 A/\epsilon_1}}$. As shown in Fig.~\ref{fig:Gaps}, self-cavity effects can be leveraged to explore the fractal dispersion relation of strange plasmons. One interesting consequence of scale invariance, is that changing the size of the screened region will re-scale resonance frequencies of the plasmons but the number of visible modes remains the same and is largely fixed by the fractal parameter $\alpha$. In this example we chose $\alpha = 0.7$ which leads to the presence of three visible resonances. This consequence of self-similar conductivity is elaborated in Suppl. Note 5\cite{Supplemental}.

Finally, another common quasiparticle found in 2D systems due to their reduced screening are excitons. While signatures of excitons have been observed at low frequencies in counterflow measurements\cite{zhang2025excitons}, probing their formation and interaction remains challenging. To demonstrate the unique insight gained from the cavity conductivity of excitons, we generate an input spectrum using absorption measurements of bilayer graphene taken at higher frequencies\cite{ju2017tunable}, shown in Fig.\ref{fig:Gaps}, top. In the cavity conductivity, the excitonic response directly demonstrates the formation of exciton-polaritons, and the hybridization between two modes which slightly shifts their resonances in the spectrum. The shifting in frequency demonstrates the dressing of light due to the self-cavity formation. 

We emphasize here that the rich phenomenology of all four examples, including the effect of edges, polaritonic dressing, and hybridization of different IR active modes are accounted for in our analytical formula which allows a direct mapping between bare and cavity conductivity. Details for the superconductivity model, the correlated insulating state, the strange metal plasmons and the hybrid exciton polaritons are found in Supp. Note~5. 
\label{sec:CorrGaps}
\paragraph{Discussion and outlook.} 
\label{sec:disc}
The analytical theory presented in this article opens the door for on-chip THz spectroscopy to become a workhorse in the study of vdW heterostructures. We have demonstrated that using this method we can extract the bare conductivity of 2D quantum materials, study finite momentum polaritonic light-matter hybrid modes as well as probe optically silent phenomena with nanometer to \textup{\AA} sensitivity along the z-direction. As mentioned in the introduction, due to limitations of different THz spectroscopies of vdW heterostructures, the combined analytical theory - experimental set-up method represents the first known approach for extracting the bare conductivity of 2D systems. 

In future studies, the in-depth understanding provided by the present analysis may enable the deliberate tuning of the cavity THz response in heterostructures. An intriguing application lies in merging cavity QED with vdW heterostructures, where photon and polariton modes can be utilized to modify the ground state of many-body systems\cite{Schlawin22,eckhardt2024,emil2025}. Our framework can facilitate the design of tailored plasmonic self-cavities capable of enhancing or suppressing specific interaction pathways among the constituent electrons.  


\subsection*{Funding}

M.H.M., H.M.B. and M.W.D. acknowledge support from the Alexander von Humboldt Foundation. H.M.B. acknowledges financial support from the European Union under the Marie Sklodowska-Curie Grant Agreement no. 101062921 (Twist-TOC). G.K. acknowledges support by the German Research Foundation through the Cluster of Excellence CUI: Advanced Imaging of Matter (EXC 2056, project ID 390715994). We acknowledge support by the Deutsche Forschungsgemeinschaft (DFG, German Research Foundation) - 508440990 and  531215165 (Research Unit ‘OPTIMAL’). We acknowledge support from the Max Planck-New York Center for Non-Equilibrium Quantum Phenomena. This research was developed with funding from the Defense Advanced Research Projects Agency (DARPA) under the QUAMELEON Advanced Research Concept. The views, opinions and/or findings expressed are those of the authors and should not be interpreted as representing the official views or policies of the Department of Defense or the U.S. Government. 

\subsection*{Acknowledgements}
We acknowledge fruitful discussions with Angel Rubio, Eugene Demler, Michael Sentef, Dante Kennes and Emmanuel Baudin.  

\subsection*{Competing interests} 
The authors declare no competing interests.

\subsection{Data, materials and code availability}

Data available upon request. For details of the analytical method while it is under review contact \textcolor{blue}{mariosm@pks.mpg.de}.

\subsection*{Author contributions}
M. H. M. developed the analytical framework and the code with support of G. K. . G. K. and M. H. M. performed the calculations with support of A. M. P. . J. W. M. and H. M. B. designed and supervised the experimental protocol. G. K. and H. M. B. carried out on-chip THz spectroscopy experiments, with the help of M. W. D. . T. M. performed numerical CST simulations with the aid of G. M.. All authors contributed in analyzing the results and writing the paper. J. W. M. supervised the overall project.

\bibliography{references}
\end{document}